\documentclass[twocolumn,unsortedaddress,showpacs,showkeys]{revtex4}
\usepackage{graphicx,psfrag,amsmath,amssymb}
\usepackage{dcolumn}
\usepackage{bm}

\begin{document}

\title{Reconstruction of stochastic nonlinear dynamical models \\ from trajectory measurements}

\author{V.~N.~Smelyanskiy$^{1}$}
\email{Vadim.N.Smelyanskiy@nasa.gov}
\author{D.~G.~Luchinsky$^{1,2}$}
\author{D.~A.~Timu{\c c}in$^{1}$}
\author{A.~Bandrivskyy$^{2}$}

\affiliation{$^{1}$NASA Ames Research Center, Mail Stop 269-2, Moffett Field, CA 94035, USA}

\affiliation{$^{2}$Department of Physics, Lancaster University, Lancaster LA1 4YB, UK}

\begin{abstract}
    A new algorithm is presented for reconstructing stochastic nonlinear dynamical models from noisy
    time-series data.  The approach is analytical; consequently, the resulting algorithm does {\em
    not} require an extensive global search for the model parameters, provides optimal compensation
    for the effects of dynamical noise, and is robust for a broad range of dynamical models.  The
    strengths of the algorithm are illustrated by inferring the parameters of the stochastic Lorenz
    system and comparing the results with those of earlier research.  The efficiency and accuracy of
    the algorithm are further demonstrated by inferring a model for a system of five globally- and
    locally-coupled noisy oscillators.
\end{abstract}



\maketitle

\section{Introduction}
\label{s:intro}
Stochastic nonlinear dynamical models are widely used in studying complex (natural as well as
man-made) phenomena; examples range from molecular motors~\cite{Visscher:99} and semiconductor
lasers~\cite{Willemsen:00a} to epidemiology~\cite{Earn:00} and coupled matter--radiation systems in
astrophysics~\cite{Christensen:02}.  Accordingly, much attention has been paid in the statistical
physics community to the central problem of reconstructing (i.e., inferring) stochastic nonlinear
dynamical models from noisy measurements (see, e.g.,~\cite{Congdon:01,Dagostini:99}).  The chief
difficulty here stems from the fact that, in a great number of important problems, it is not
possible to derive a suitable model from ``first principles,'' and one is therefore faced with a
rather broad range of possible parametric models to consider.  Furthermore, experimental data can
sometimes be extremely skewed due to the intricate interplay between noise and nonlinearity, making
it very difficult to extract from data important ``hidden'' features (e.g., coupling parameters) of
a model.

Although no general method exists for inferring the parameters of stochastic nonlinear dynamical
models from measurements, various schemes have been proposed
recently~\cite{McSharry:99a,Heald:00,Meyer:00a,Meyer:01,Rossi:02a,Friedrich:98,Friedrich:00,Friedrich:00b,Friedrich:03a}
to deal with different aspects of this ``inversion'' problem.  An important numerical technique,
suggested in~\cite{Friedrich:98,Friedrich:00,Friedrich:00b}, is based on estimating drift and
diffusion coefficients at a number of points in the phase space of the dynamical system.  This
technique was extended further in~\cite{Friedrich:03a} to handle both dynamical and measurement
noise.  In principle, this approach allows subsequent use of the least-squares method for the
estimation of the model parameters.  Such an empirical approach, however, requires a considerable
amount of data and an intensive computational effort even for a simple stochastic equation.  A more
general, and efficient, theoretical approach is therefore very desirable.

Arguably the most general approach to the solution of this problem is furnished by Bayes'
theorem~\cite{Heald:00,Meyer:01,Bremer:01}.  Indeed, it was shown in~\cite{Meyer:00a} that the
Bayesian method provides a rigorous and systematic basis for heuristic
modifications made earlier~\cite{McSharry:99a} to the least-squares method to enable its use on
noisy measurements.  The Bayesian approach was employed in~\cite{Heald:00} to estimate levels of
dynamical and measurement noise for a known dynamical model.  The Bayesian method has also been used
for parameter estimation in maps in the presence of dynamical~\cite{Meyer:00a} and weak
measurement~\cite{Meyer:01} noise.  Finally, an application of the Bayesian method to continuous
systems was considered in~\cite{Rossi:02a}.

A common drawback of these earlier works is their exclusive reliance on numerical methods for the
optimization of cost functions and the evaluation of multi-dimensional normalization integrals
encountered in the theory.  This disadvantage becomes increasingly more pronounced when systems
with ever larger numbers of unknown parameters are investigated.  Another major deficiency is that
most of the earlier works deal with discrete maps, and the corresponding results are therefore not
immediately applicable to continuous systems, since the transformation from noise variables to
dynamical variables is different in discrete and continuous cases.  Specifically, as will be shown
below, a prefactor accounting for the Jacobian of the transformation must be included in the
likelihood function in the continuous case.  Such a prefactor was considered in~\cite{Rossi:02a} in
the context of Bayesian inference for continuous systems; however, an {\it ad hoc} likelihood
function was used there instead of the correct form derived here.

In this paper, we introduce a new technique for Bayesian inference of stochastic nonlinear dynamical
models from noisy measurements.  At the core of our algorithm is a path-integral representation of
the likelihood function that yields the correct form for the Jacobian prefactor.  This term provides
optimal compensation for the effects of dynamical noise, thus leading to robust inference for a
broad range of dynamical models.  Another key feature of the approach is a novel parameterization of
the vector ``force'' field, which permits an analytical treatment of the inference problem, thus
obviating the need for extensive global optimization.  These improvements lead to an efficient and
accurate algorithm for reconstructing from time-series data models of stochastic nonlinear dynamical
systems with large numbers of unknown parameters.

The paper is organized as follows.  The general formulation of the problem and its analytical
solution are presented in Section~\ref{s:main}.  The algorithm is then applied in
Section~\ref{ss:Lorenz} to data from the stochastic Lorenz system, and its performance is compared
with those of earlier research.  The advantages of the present method are further illustrated in
Section~\ref{ss:five_coupled} by inferring a model for a system of five globally- and
locally-coupled noisy oscillators.  Finally, the results are discussed and conclusions are drawn in
Section~\ref{s:discussion}.

\section{Theory of reconstruction of stochastic nonlinear dynamical models}
\label{s:main}
\subsection{Problem description}
\label{ss:problem}
We envision a typical experimental situation where the stochastic trajectory ${\bf x}(t)$ of a
dynamical system is measured at sequential time instants $\lbrace t_n; n = 0, 1, \ldots, N \rbrace$,
and a set of data ${\cal Y} = \lbrace {\bf y}_n \equiv {\bf y}(t_{n}) \rbrace$ is thus obtained.
For instance, ${\bf x}(t)$ may represent the coordinates of a molecular motor progressing along a
microtubule~\cite{Visscher:99} or the fluctuating Stokes vector of a semiconductor laser
field~\cite{Willemsen:00a}.  Our objective is to extract from ${\cal Y}$ all available information
regarding the dynamical evolution of ${\bf x}(t)$.  As mentioned in Section~\ref{s:intro}, we
advocate the Bayesian approach for the solution of this problem.  Toward this end, one has to
introduce a parametric model for the dynamical system and a statistical model for the measurements.
These elements allow one to incorporate into the solution of the reconstruction problem any
available {\it a priori} information on the time-series data (stationarity, embedding dimension,
etc.), as well as expert domain knowledge (e.g., a theoretical analysis of the physics problem at
hand).

The dynamical and measurement equations commonly adopted in the context of model reconstruction are
\begin{equation}
    \left.
    \begin{array}{rcl}
    \dot{\bf x}(t) & = & {\bf f}({\bf x}; {\bf c}) + {\boldsymbol \xi}(t), \\
    {\bf y}(t) & = & {\bf x}(t) + {\boldsymbol \nu}(t), \\
    \end{array}
    \quad \right\}
    \label{eq:dynamics}
\end{equation}
with ${\bf x}, {\bf y}, {\boldsymbol \xi}, {\boldsymbol \nu} \in {\mathbb R}^L$, and ${\bf f}:
{\mathbb R}^L \mapsto {\mathbb R}^L$.  Here, the first equation represents the dynamical model in
the form of a set of coupled nonlinear Langevin equations with a vector field ${\bf f}({\bf x}; {\bf
c})$ parameterized by unknown coefficients ${\bf c} \in {\mathbb R}^M$, and the second equation
relates the observations to the system trajectory.  We assume that the additive dynamical and
measurement noise processes ${\boldsymbol \xi}(t)$ and ${\boldsymbol \nu}(t)$ are stationary, white,
and Gaussian with
\begin{equation}
    \left.
    \begin{array}{rcl}
        \langle {\boldsymbol \xi}(t) \rangle = {\bf 0}, & \quad & \hspace{0.03in} \langle
        {\boldsymbol \xi}(t) \, {\boldsymbol \xi}^{\rm T}(t') \rangle = \hat{\bf D} \,
        \delta(t - t'), \\
    \langle {\boldsymbol \nu}(t) \rangle = {\bf 0}, & \quad & \langle {\boldsymbol \nu}(t) \,
    {\boldsymbol \nu}^{\rm T}(t') \rangle = \epsilon^{2} \, \hat{\bf I} \, \delta(t - t'),
    \end{array}
    \quad \right\}
    \label{eq:noise}
\end{equation}
where $\hat{\bf D}$ and $\epsilon$ are also typically unknown.  Thus, the elements $\lbrace c_m; m =
1, 2, \ldots, M \rbrace$ of the model coefficient vector ${\bf c}$, the elements $\lbrace D_{l l'};
l,l' = 1, 2, \ldots, L \rbrace$ of the dynamical noise covariance (or diffusion) matrix $\hat{\bf
D}$, and the measurement noise intensity $\epsilon^2$ together constitute the complete set
\begin{equation}
    \label{eq:unknowns}
    {\cal M} = \lbrace {\bf c}, \hat{\bf D}, \epsilon \rbrace
\end{equation}
of unknown parameters.  The model reconstruction problem, then, is that of inferring the elements of
the parameter set ${\cal M}$ from the measured time-series data ${\cal Y}$.

\subsection{Bayesian inference}
\label{ss:inference}
In Bayesian model inference, two distinct probability density functions (PDFs) are ascribed to the
set of unknown model parameters: the {\em prior} $p_{\textrm{pr}}({\cal M})$ and the {\em posterior}
$p_{\textrm{ps}}({\cal M}|{\cal Y})$, respectively representing our state of knowledge about
${\cal M}$ before and after processing a block of data ${\cal Y}$.  These two PDFs are related to
each other via Bayes' theorem~\cite{Congdon:01}:
\begin{equation}
    \label{eq:Bayes}
    p_{\textrm{ps}}({\cal M}|{\cal Y}) = \frac{p({\cal Y}|{\cal M}) \, p_{\textrm{pr}}({\cal
    M})}{\int p({\cal Y}|{\cal M}) \, p_{\textrm{pr}}({\cal M}) \, {\rm d}{\cal M}}.
\end{equation}
Here, the {\em sampling distribution} $p({\cal Y}|{\cal M})$ is the conditional PDF of the
measurements ${\cal Y}$ for a given choice of the model ${\cal M}$; it is also referred to, as we
do, as the {\em likelihood} of ${\cal M}$ given ${\cal Y}$.  Meanwhile, the prior acts as a {\em
regularizer}, concentrating the parameter search to those regions of the model space favored by our
expertise and any available auxiliary information.  This initial assignment of probabilities must,
of course, be ``coherent''~\cite{Jeffreys:61}, i.e., consistent, at least implicitly, with the
physics of the problem.  In practice, (\ref{eq:Bayes}) can be applied iteratively using a sequence
of data blocks ${\cal Y}, {\cal Y}^{\prime}, \ldots$; the posterior computed from block ${\cal Y}$
serves as the prior for the next block ${\cal Y}^{\prime}$, etc.  For a sufficiently large number of
observations, $p_{\textrm{ps}}({\cal M}|{\cal Y}, {\cal Y}^{\prime}, \ldots)$ becomes sharply peaked
around a most probable model ${\cal M}^{\ast}$.

\begin{figure}[h]
   \begin{center}
 \includegraphics[width=6cm,height=5cm]{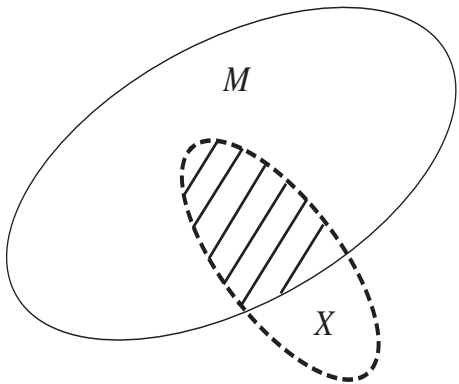}
   \end{center}
\caption{\label{fig:bayes} Illustration of the geometrical
interpretation and physical meaning the Bayes theorem. ${\cal M}$
denotes set of all possible values of the model parameters. Events
corresponding to the measured time-series data are shown by ${\cal
X}$. Intersection of the two sets is shown by the shaded area. The
geometrical interpretation of the Bayes theorem is that the
likelihood function cuts out of the all possible model parameters
those that correspond to the observed time-series data.}
 \end{figure}

We note that Bayes theorem allow for a simple geometric
interpretation and has a clear physical sense. Indeed, denote by
${\cal M}$ all the events corresponding every possible values of
the model parameters (these events are encircled in the figure
\ref{fig:bayes} by the solid line).
The events corresponding to the measured time-series data are
shown by ${\cal X}$. Then the intersection of the two sets of
events can be factories in two ways $P({\cal X}\bigcap{\cal
M})$=$P({\cal X}|{\cal M})P({\cal M})$=$P({\cal M}|{\cal
X})P({\cal X})$, where $P({\cal X}|{\cal M})$ and $P({\cal
M}|{\cal X})$ are corresponding conditional probabilities. Taking
into account that $P({\cal X})=\int d{\cal M}P({\cal X}|{\cal
M})P({\cal M})$ and dividing last equation by $P({\cal X})$ we
have Bayes theorem. then geometrical interpretation of the Bayes
theorem is that the likelihood function cuts out of the all
possible model parameters those that correspond to the observed
time-series data. It has a clear physical meaning that states that
given initial guess about model parameters it can be improved
using results of the measurements. This theorem plays fundamental
role in the modern theory of measurements.


The main thrust of recent research on stochastic nonlinear
dynamical model
reconstruction~\cite{McSharry:99a,Meyer:00a,Meyer:01,Rossi:02a}
has been directed towards developing (i) efficient optimization
algorithms for extracting the most probable model ${\cal
M}^{\ast}$ from the posterior, and (ii) efficient
multi-dimensional integration techniques for evaluating the
normalization factor in the denominator of (\ref{eq:Bayes}).
These efforts have mostly employed {\it ad hoc} expressions for
the likelihood function (see, e.g., the cost function of Eq.  (31)
in~\cite{Rossi:02a}); consequently, the resulting inference
schemes fail to properly compensate for the effects of noise.  In
fact, it appears that there is a lack in the model-reconstruction
literature of a closed-form expression (expanded to correct orders
in the sampling period) for the likelihood function of the
measurements of a continuous system trajectory.

Below we introduce a new approach to Bayesian inference of stochastic nonlinear dynamical models.
The method has two key analytical features.  Firstly, the likelihood function is written in the form
of a path integral over the stochastic system trajectory, which includes a prefactor that optimally
compensates for the detrimental effects of (dynamical) noise.  Secondly, we suggest a novel
parameterization of the unknown vector field, which renders the inference problem essentially linear
for a broad class of nonlinear dynamical systems, and thus helps us find optimal parameter estimates
without extensive numerical optimization.  These features enable us to write an efficient and
accurate Bayesian inference algorithm for reconstructing models of nonlinear dynamical systems
driven by noise.  As a prelude to the formal development that follows, the reader may at this point
wish to review the theory given in Appendix A for the maximum-likelihood reconstruction of a
one-dimensional system model.

\subsection{The likelihood function}
\label{ss:likelihood}
As we pointed out above, one of the central challenges in the inference of stochastic nonlinear
dynamical models is the derivation of a suitable likelihood function that optimally compensates for
the effects of noise.  A key ingredient in this context is the probability density functional ${\cal
F}_{\cal M}[{\bf x}(t)]$ of finding the system in ``state'' ${\bf x}(t)$ at time
$t$~\cite{Rossi:02a,Borland:92a,Borland:92b,Borland:96}.  This is supplemented by $p_{\rm ob}({\cal
Y}|{\cal X})$ denoting the PDF of observing a time series ${\cal Y}$ for a specific realization
${\cal X} = \{{\bf x}_n\}$ of the system trajectory.  Thus, we may express the likelihood function
very generally in the form of a path integral over the random trajectories of the dynamical system
as
\begin{equation}
    \label{eq:pathint}
    p({\cal Y}|{\cal M}) = \int_{{\bf x}(t_{\rm i})}^{{\bf x}(t_{\rm f})} p_{\rm ob}({\cal
    Y}|{\cal X}) \, {\cal F}_{\cal M}[{\bf x}(t)] \, {\cal D}{\bf x}(t),
\end{equation}
giving the probabilistic relationship between the observations ${\cal Y}$ and the unknown parameters
${\cal M}$ of the model (\ref{eq:dynamics}).  Here, we choose $t_{\rm i} \ll t_{0} < t_{N} \leqslant
t_{\rm f}$ so that $p({\cal Y}|{\cal M})$ does not depend on the particular initial and final states
${\bf x}(t_{\rm i})$, ${\bf x}(t_{\rm f})$.  We note that the path-integral approach has also proved
useful in nonlinear filtering of random signals (see, e.g.,~\cite{Rosov:02}) where standard spectral
and correlation analyses fail.

The explicit form of ${\cal F}_{\cal M}[{\bf x}(t)]$ has been given
in~\cite{Graham:77a,Gozzi:83,Dykman:90}; however, in the context of dynamical inference, it is not
necessary to employ this exact form as one can usually rely on the smallness of the sampling
interval.  Accordingly, adopting a uniform sampling scheme $t_{n} = t_{0} + n h$, we assume here for
the sake of simplicity that $h \equiv (t_{N} - t_{0})/N$ is small, and rewrite (\ref{eq:dynamics})
using a mid-point Euler discretization scheme in the form
\begin{equation}
    \left.
    \begin{array}{rcl}
    {\bf x}_{n+1} & = & {\bf x}_{n} + h \, {\bf f}(\tilde{\bf x}_{n}; {\bf c}) + {\bf z}_n, \\
    {\bf y}_n & = & {\bf x}_n + {\boldsymbol \nu}_n,
    \end{array}
    \quad \right\}
    \label{eq:discrete_dynamics}
\end{equation}
where $\tilde{\bf x}_{n} \equiv \frac{1}{2} \, ({\bf x}_{n+1} + {\bf x}_n)$, while ${\bf z}_n$ are
independent, zero-mean, Gaussian random variables with covariance $\langle {\bf z}_{n} \, {\bf
z}_{n'}^{\rm T} \rangle = h \, \hat{\bf D} \, \delta_{n n'}$.  The probability of a particular
realization $\{{\bf z}_n\}$ of the dynamical noise process is simply
\begin{eqnarray}
    \label{eq:markov}
    {\cal P}[\{{\bf z}_n\}] = \prod_{n=0}^{N-1} \frac{{\mathrm d}{\bf z}_{n}}{\sqrt{(2 \pi h)^{L}
    |\hat{\bf D}|}} \, \exp \left( -\frac{1}{2 h} \, {\bf z}_n^{\mathrm T} \, {\hat{\bf D}}^{-1} \,
    {\bf z}_n \right).
\end{eqnarray}
Changing to dynamical state variables using (\ref{eq:discrete_dynamics}), we thus obtain the desired
PDF for the dynamical system (\ref{eq:dynamics}) to have an arbitrary trajectory $\{{\bf x}_n\}$:
\begin{widetext}
    \begin{eqnarray}
    \label{eq:markov_dynamic}
    {\cal F}_{\cal M}[\{{\bf x}_n\}] & = & p_{\rm st}({\bf x}_{0}) \, J(\{{\bf x}_{n}\}) \nonumber
    \\
    & \times & \prod_{n=0}^{N-1} \frac{1}{\sqrt{(2 \pi h)^{L} |{\hat {\bf D}}|}} \, \exp
    \left(-\frac{1}{2 h} \, [{\bf x}_{n+1} - {\bf x}_{n} - h \, {\bf f}(\tilde{\bf x}_{n}; {\bf
    c})]^{\rm T} \, {\hat {\bf D}}^{-1} \, [{\bf x}_{n+1} - {\bf x}_{n} - h \, {\bf f}(\tilde{\bf
    x}_{n}; {\bf c})] \right),
    \end{eqnarray}
\end{widetext}
where $p_{\rm st}({\bf x})$ signifies the stationary distribution of ${\bf x}(t)$, and the Jacobian
of the transformation is given by
\begin{eqnarray}
    \label{eq:prefactor}
    J(\{{\bf x}_n\}) & = & \left| \left\{ \frac{\partial z_{l n}}{\partial x_{l' n'}} \right\}
    \right| \simeq \prod_{n=1}^{N} \prod_{l=1}^{L} \left[ 1 - \frac{h}{2} \, \frac{\partial
    f_{l}(\tilde{\bf x}_{n}; {\bf c})}{\partial x_{l n}} \right] \nonumber \\
    & \simeq & \exp \left[-\frac{h}{2} \, \sum_{n=1}^{N} \mathop{\rm{tr}} {\hat {\bf
    \Phi}}(\tilde{\bf x}_{n}; {\bf c}) \right],
\end{eqnarray}
approximated to leading order in $h$, with the elements of the matrix ${\hat {\bf \Phi}}$ defined as
$\Phi_{l l'}({\bf x}; {\bf c}) \equiv \partial f_{l}({\bf x}; {\bf c})/\partial x_{l'}$.

The evaluation of (\ref{eq:pathint}) requires, in addition, that we adopt a specific form for the
measurement PDF $p_{\rm ob}({\cal Y}|{\cal X})$.  We assume here that, for each trajectory component
$x_l(t)$, the measurement error $\epsilon$ is negligible compared with the fluctuations induced by
the dynamical noise; i.e., $\epsilon^2 \ll h D_{l l}$.  Consequently, we may use
\begin{equation}
    \label{eq:observer}
    p_{\rm ob}({\cal Y}|{\cal X}) \simeq \prod_{n = 0}^{N} \delta({\bf y}_{n} - {\bf x}_n)
\end{equation}
in (\ref{eq:pathint}), and the set of unknown model parameters to be inferred from data reduces to
${\cal M} = \lbrace {\bf c}, {\hat {\bf D}} \rbrace$.  With this substitution, (\ref{eq:pathint}) is
easily evaluated; introducing $\tilde{\bf y}_{n} \equiv \frac{1}{2} \, ({\bf y}_{n+1} + {\bf y}_n)$,
we write the result in the form
\begin{widetext}
    \begin{eqnarray}
    \label{eq:likelihood}
    -\frac{2}{N} \, \ln p({\cal Y}|{\cal M}) & = & -\frac{2}{N} \, p_{\rm st}({\bf y}_{0}) + L \ln (2
    \pi h) + \ln |{\hat{\bf D}}| \nonumber \\
    & + & \frac{h}{N}\, \sum_{n=0}^{N-1} \left\{ \mathop{\rm{tr}} {\hat {\bf \Phi}}(\tilde{\bf
    y}_{n}; {\bf c}) + [\dot{\bf y}_{n} - {\bf f}(\tilde{\bf y}_{n}; {\bf c})]^{\rm T} \, {\hat {\bf
    D}}^{-1} \, [\dot{\bf y}_{n} - {\bf f}(\tilde{\bf y}_{n}; {\bf c})] \right\},
    \end{eqnarray}
\end{widetext}
where we introduced the ``velocity'' $\dot {\bf y}_{n} \equiv ({\bf y}_{n+1} - {\bf y}_n)/h$.  It is
important to note that this likelihood function is asymptotically exact in the limit $h \rightarrow
0$ and $N \rightarrow \infty$, with the total observation duration $T = N h$ remaining constant.

It is the term $\mathop{\rm{tr}}{\hat {\bf \Phi}}(\tilde{\bf
y}_{n}; {\bf c})$ in the above that provides optimal compensation
for the detrimental effects of dynamical noise, and distinguishes
our likelihood function from those introduced in earlier works.
Formally, this term emerges from the path integral as the Jacobian
of the transformation from noise variables to dynamical
variables~\cite{Graham:73,Gozzi:83}.  We emphasize, however, that
this is {\em not} merely a correction term, but is in fact crucial
for accurate inference. In particular, for a small attractor (with
characteristic length scale smaller then square root of the noise
intensity) the inference is only possible due to this term as will
be shown in Section~\ref{s:models}.

\subsection{Parameterization of the unknown vector field}
\label{ss:parameterization}
As mentioned in Section~\ref{s:intro}, one of the main difficulties encountered in the inference of
stochastic nonlinear dynamical models is that the cost function, defined in (\ref{eq:action}) below,
is generally nonlinear in the model parameters, thus requiring the use of extensive numerical
optimization methods for finding its global minimum.  The parameterization we now introduce avoids
this difficulty while still encompassing a broad class of nonlinear dynamical models.  Indeed, many
of the model reconstruction examples considered in earlier works on stochastic nonlinear dynamical
inference can be solved within this framework.  Moreover, a large number of important practical
applications (see, e.g.,~\cite{Rossi:97,Nigmatulin:00}) can also be treated using the same approach.

We parameterize the nonlinear vector field in the form
\begin{equation}
    \label{eq:model}
    {\bf f}({\bf x}; {\bf c}) = {\hat {\bf U}}({\bf x}) \, {\bf c},
\end{equation}
where ${\hat {\bf U}}({\bf x})$ is an $L \times M$ matrix of suitably chosen basis functions, and
${\bf c}$ is an $M$-dimensional vector of unknown parameters.  The choice of basis functions is open
to any appropriate class of (polynomial, trigonometric, etc.)  functions that may be required for a
satisfactory representation of the vector field.  In general, if we use $G$ different basis
functions $\lbrace \phi_g({\bf x}); \, g = 1, 2, \ldots, G \rbrace$ to model the vector field ${\bf
f}$, then the matrix $\hat{\bf U}$ will have the block structure
\begin{widetext}
    \begin{equation}
    \label{eq:block_structure_1}
    \hat{\bf U} = \left[ \left(
    \begin{array}{cccc}
        \phi_1 & 0 & \ldots & 0 \\
        0 & \phi_1 & \ldots & 0 \\
        \vdots & \vdots & \ddots & \vdots \\
        0 & 0 & \ldots & \phi_1 \\
    \end{array} \right) \,
    \left(
    \begin{array}{cccc}
        \phi_2 & 0 & \ldots & 0 \\
        0 & \phi_2 & \ldots & 0 \\
        \vdots & \vdots & \ddots & \vdots \\
        0 & 0 & \ldots & \phi_2 \\
    \end{array} \right)
    \ldots \left(
    \begin{array}{cccc}
        \phi_G & 0 & \ldots & 0 \\
        0 & \phi_G & \ldots & 0 \\
        \vdots & \vdots & \ddots & \vdots \\
        0 & 0 & \ldots & \phi_G \\
    \end{array} \right)
    \right],
    \end{equation}
\end{widetext}
comprising $G$ diagonal blocks of size $L \times L$ ($M = G L$), with the $\bf x$ dependence
suppressed for brevity.  An important feature of (\ref{eq:model}) for our subsequent development is
that, while possibly highly nonlinear in ${\bf x}$, ${\bf f}({\bf x}; {\bf c})$ is strictly linear
in ${\bf c}$.

As shown next, (\ref{eq:likelihood}) and (\ref{eq:model}) are the two main ingredients that enable
an analytic solution to the problem of stochastic nonlinear dynamical model inference from
time-series data.

\subsection{The algorithm}
\label{ss:algorithm}
We start by choosing a prior model PDF that is Gaussian in $\bf c$ and uniform in $\hat{\bf D}$:
\begin{equation}
    p_{\textrm{pr}}({\cal M}) \propto \exp \left[-\frac{1}{2} ({\bf c}-{\bf
    c}_{\textrm{pr}})^{\mathrm T} \, \hat{\bf \Sigma}_{\textrm{pr}} \, ({\bf c}-{\bf
    c}_{\textrm{pr}}) \right].
    \label{eq:prior}
\end{equation}
Substituting (\ref{eq:model}), (\ref{eq:prior}), and the likelihood $p({\cal Y}|{\cal M})$ given by
(\ref{eq:likelihood}) into (\ref{eq:Bayes}), we obtain the posterior model PDF in the form
$p_{\textrm{ps}}({\cal M}|{\cal Y}) = {\rm const} \times \exp[-S_{\cal{Y}}({\bf c}, \hat{\bf D})]$,
where
\begin{equation}
    S_{\cal{Y}}({\bf c}, \hat{\bf D}) = \rho_{\cal{Y}}(\hat{\bf D}) - {\bf
    c}^{\mathrm T} \, {\bf w}_{\cal{Y}}(\hat{\bf D}) + \frac{1}{2} \, {\bf c}^{\mathrm T} \,
    \hat{\bf \Xi}_{\cal{Y}}(\hat{\bf D}) \, {\bf c}
    \label{eq:action}
\end{equation}
is the cost function whose global minimum yields the most probable model ${\cal M}^{*} = \lbrace {\bf
c}^{*}, {\hat {\bf D}^{*}} \rbrace$.  Here, use was made of the definitions
\begin{eqnarray}
    \label{eq:defs_1}
    \hspace{-0.25in} \rho_{\cal Y}(\hat{\bf D}) & = & \frac{h}{2} \, \sum_{n = 0}^{N - 1} \dot{{\bf
    y}}_{n}^{\mathrm T} \, \hat{\bf D}^{-1} \, \dot{{\bf y}}_{n} + \frac{N}{2} \, \ln |\hat{\bf D}|,
    \\
    \label{eq:defs_2}
    \hspace{-0.25in} {\bf w}_{\cal Y}(\hat{\bf D}) & = & \hat{\bf \Sigma}_{\textrm{pr}} \, {\bf
    c}_{\textrm{pr}} + h \, \sum_{n = 0}^{N - 1} \left[ \hat{\bf U}_{n}^{\mathrm T} \, \hat{\bf
    D}^{-1} \, \dot{{\bf y}}_{n}- \frac{1}{2} \, {\bf v}_{n} \right], \\
    \label{eq:defs_3}
    \hspace{-0.25in} \hat{\bf \Xi}_{\cal Y}(\hat{\bf D}) & = & \hat{\bf \Sigma}_{\textrm{pr}} + h
    \, \sum_{n = 0}^{N - 1} \hat{\bf U}_{n}^{\mathrm T} \, \hat{\bf D}^{-1} \, \hat{\bf U}_{n},
\end{eqnarray}
where $\hat{\bf U}_{n} \equiv \hat{\bf U}(\tilde{\bf y}_{n})$, ${\bf v}_{n} \equiv {\bf
v}(\tilde{\bf y}_{n})$, and the components of the vector ${\bf v}({\bf x})$ are
\begin{equation}
    v_{m}({\bf x}) = \sum_{l = 1}^{L} \frac{\partial U_{lm}({\bf x})}{\partial x_l}, \quad m = 1, 2,
    \ldots, M.
    \label{eq:v}
\end{equation}

For a given block of data ${\cal Y}$ of length $(N + 1)$, the best estimates for the model
parameters are given by the posterior means of ${\bf c}$ and $\hat{\bf D}$, which coincide with the
global minimum of $S_{\cal Y}({\bf c}, \hat{\bf D})$.  We handle this optimization problem in the
following way.  Assume for the moment that ${\bf c}$ is known in (\ref{eq:action}); for the first
iteration, take ${\bf c} = {\bf c}_{\textrm{pr}}$.  Then, minimizing $S_{\cal Y}({\bf c}, \hat{\bf
D})$ with respect to $\hat{\bf D}$, we find that the posterior distribution over $\hat{\bf D}$ has
a mean
\begin{equation}
    \label{eq:updateD}
    \langle \hat{\bf D} \rangle = \frac{1}{N} \, \sum_{n=0}^{N-1} \left( {\dot {\bf y}}_{n} -
    \hat{\bf U}_{n} \, {\bf c} \right) \left( {\dot {\bf y}}_{n} - \hat{\bf U}_{n} \, {\bf c}
    \right)^{\mathrm T}.
\end{equation}
Assume next that $\hat{\bf D}$ is known, and note from (\ref{eq:action}) that in this case, the
posterior distribution over ${\bf c}$ is Gaussian.  Its covariance is given by $\hat{\bf \Xi}_{\cal
Y}(\hat{\bf D})$, and its mean
\begin{equation}
    \langle {\bf c} \rangle = \hat{\bf \Xi}^{-1}_{\cal Y}(\hat{\bf D}) \, {\bf w}_{\cal Y}(\hat{\bf
    D})
    \label{eq:updateC}
\end{equation}
minimizes $S_{\cal Y}({\bf c}, \hat{\bf D})$ with respect to ${\bf c}$.  Thus, for the second
iteration, ${\bf c}_{\textrm{pr}}$ and $\hat{\bf \Sigma}_{\textrm{pr}}$ are replaced with $\langle
{\bf c} \rangle$ and $\hat{\bf \Xi}_{\cal Y}(\hat{\bf D})$, respectively.  This two-step
(analytical) optimization procedure is continued iteratively until convergence, which is typically
much faster than brute-force numerical optimization that has been attempted in earlier works.

It is worthwhile to pause here and reflect on the content of (\ref{eq:defs_2}).  The first term in
the sum is essentially the generalized least-squares (GLS) result (see Appendix B), and vanishes at
the attractors of the dynamical system (\ref{eq:dynamics}).  On the other hand, the second term in
the sum on the right-hand side of (\ref{eq:defs_2}), originating from the term ${\rm tr} \, \hat{\bf
\Phi}(\tilde{\bf y}_{n}; {\bf c})$ in (\ref{eq:likelihood}), does not vanish at an attractor, and is
in fact crucial for accurate inference in the presence of noise.  This can be demonstrated
analytically by rewriting the sum in integral form as
\begin{eqnarray}
    \label{eq:integral}
    {\bf w}_{\cal Y}(\hat{\bf D}) & = & \hat{\bf \Sigma}_{\textrm{pr}} \, {\bf c}_{\textrm{pr}} +
    \int_{{\bf x}(t_0)}^{{\bf x}(t_{0} + T)} \hat{\bf U}[{\bf y}(t)]^{\rm T} \, \hat{\bf D}^{-1}
    \, {\mathrm d}{\bf y} \nonumber \\
    & - & \frac{1}{2} \, \int_{t_0}^{t_{0} + T} {\bf v}[{\bf y}(t)] \, {\mathrm d}t.
\end{eqnarray}
It can now be seen that, for an attractor localized in the phase space, the first integral will
remain finite since the initial and final points of integration both belong to the attractor.
Meanwhile, the second integral in (\ref{eq:integral}) will grow with the duration of observation
$T$.  In particular, for a point attractor, the first integral is identically zero and the second,
``compensating'' term alone contributes to inference.  This result is intuitively clear since, in
the absence of noise, the system will stay forever at the same point (i.e., the point attractor) and
no structure can be inferred.  It is the dynamical noise that forces the system to move about in the
phase space, thus making it possible to infer its structure from time-series data.

In general, then, both of the integral terms in (\ref{eq:integral}) are needed to optimally
compensate for the effects of dynamical noise and thus enable robust convergence of our inference
algorithm.  The relative importance of these two terms will be investigated quantitatively in the
following section.

\section{Inference examples}
\label{s:models}
We have verified the accuracy and robustness of our algorithm on several different types of
dynamical systems.  Here, we discuss its performance on two representative examples.

\subsection{The Lorenz system}
\label{ss:Lorenz}
\begin{figure}
    \centering
    \psfrag{x1}[cc][cc]{$x_{1}$}
    \psfrag{x2}[cc][cc]{$x_{2}$}
    \psfrag{x3}[cc][cc]{$x_{3}$}
    \psfrag{(a)}[cc][cc]{(a)}
    \psfrag{(b)}[cc][cc]{(b)}
    \includegraphics[width=3.25in]{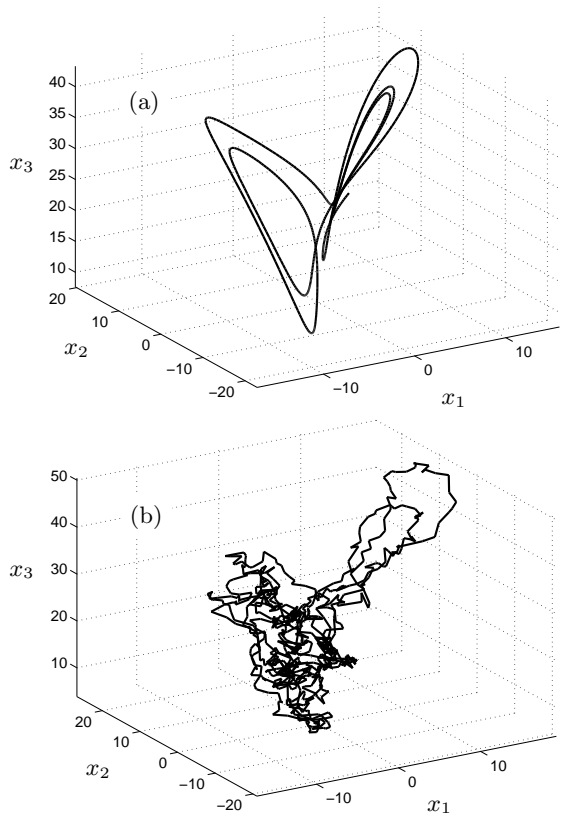}
    \caption{\label{fig:Lorenz} The phase portrait of the chaotic nonlinear Lorenz system
    (\ref{eq:Lorenz}) with the standard parameters (see text): (a)~deterministic system;
    (b)~stochastic system with strong dynamical noise, simulated with a diagonal diffusion matrix
    having elements $D_{11} = 1500$, $D_{22} = 1600$, and $D_{33} = 1700$.  (All quantities in the
    equations and figures are dimensionless in this paper.)}
\end{figure}

We start with the archetypical chaotic nonlinear system of Lorenz,
\begin{equation}
    \label{eq:Lorenz}
    \left.
    \begin{array}{rcl}
    {\dot x}_{1} & = & \sigma \, (x_{2} - x_{1}) + \xi_1(t), \\
    {\dot x}_{2} & = & r \, x_{1} - x_{2} - x_{1} \, x_{3} + \xi_2(t), \\
    {\dot x}_{3} & = & x_{1} \, x_{2} - b \, x_{3} + \xi_3(t),
    \end{array}
    \quad \right\}
\end{equation}
augmented by zero-mean Gaussian noise processes $\xi_{l}(t)$ with covariance $\langle \xi_{l}(t) \,
\xi_{l'}(t') \rangle = D_{l l'} \, \delta(t - t')$.  Synthetic data (with no measurement noise) were
generated by simulating (\ref{eq:Lorenz}) using the standard parameter set $\sigma = 10$, $r = 28$,
$b = \frac{8}{3}$, and for various levels of dynamical noise intensities as explained below.  The
phase portrait of the Lorenz system with dynamical noise is shown in Figure~\ref{fig:Lorenz} along
with the noiseless case to visually convey the difficulty of the inference problem.

\subsubsection{Parameter estimation with strong dynamical noise}

We compare now the performance of our algorithm with the results
of earlier work~\cite{Rossi:02a}. No attempt was made
in~\cite{Rossi:02a} to identify the model of the system and only
four unknown parameters were estimated.

In parameter estimation, the functional form of the nonlinear
force field -- in this case, the right-hand side of
(\ref{eq:Lorenz}) -- is assumed known, and the associated
coefficients are then estimated from data.  This is the approach
reported in~\cite{Rossi:02a}, where the diffusion matrix is taken
in the form $\hat{\bf D} = \tau^2 \, \hat{\bf I}$, and the unknown
parameters $\{\sigma, r, b, \tau^2\}$ are estimated via extensive
numerical optimization of a cost function by simulated annealing
and back-propagation techniques.  We now demonstrate that our
algorithm can estimate the parameters of the system
(\ref{eq:Lorenz}) extremely efficiently and with very high
accuracy.

First we notice that since the diffusion matrix is diagonal, our
algorithm is reduces in this case to the trivial one-dimensional
analytical solution of the problem for each equation in the form
(cf with (\ref{eq:defs_2}), (\ref{eq:defs_3}), (\ref{eq:updateD}),
(\ref{eq:updateC}) and see Appendix A for the details)
\[{\bf c}_i = {\bf\hat H}_i^{-1}{\bf w}_i, \qquad i=1\ldots 3\]
where
 \[w_{il} = \sum_{n=0}^{N-1}\left(c_{il}\phi_{il}-
 \frac{\tau^2}{2}\frac{\partial \phi_{il}}{\partial x_i}\right)\]
and
\[{\bf\hat H}_i = \sum_{n=0}^{N-1}\left(
    \begin{array}{llll}
    \phi_{i1}\phi_{i1} & \phi_{i1}\phi_{i2} & \ldots & \phi_{i1}\phi_{iL} \\
    \phi_{i2}\phi_{i1} & \phi_{i2}\phi_{i2} & \ldots & \phi_{i2}\phi_{iL} \\
    \vdots & \vdots & \ddots & \vdots \\
    \phi_{iL}\phi_{i1} & \phi_{iL}\phi_{i2} & \ldots & \phi_{iL}\phi_{iL} \\
    \end{array} \right).\]
Noise intensity is found according to (\ref{eq:updateD}). We note
that in each equation we now have different basis functions
$\phi_{il}$. For the first equation we have the following two
basis functions: $\phi_{11}=x_1$ and $\phi_{12}=x_2$. For the
second equation we have: $\phi_{21}=x_1$, $\phi_{22}=x_2$, and
$\phi_{23}=x_1x_3$. And for the last equation we have:
$\phi_{31}=x_1x_2$, $\phi_{32}=x_3$.

Thus there are a total of 8 unknown parameters to be estimated: a
seven-dimensional coefficient vector $\bf c$ and the noise
intensity $\tau^2$. (Note that this is already more ambitious than
what was done in~\cite{Rossi:02a}, since we are attempting to
estimate {\em all} model coefficients, including those that are
equal to $\pm 1$.)

The convergence of our scheme is so rapid that it is feasible to
use the algorithm in real time on ``streaming'' data. To make a
fair comparison we use the same number of data points as
in~\cite{Rossi:02a}.  As an indication of the inference accuracy,
we quote in Table~\ref{tab:Lorenz_par} results for data simulated
with the standard Lorenz parameter set and two values of dynamical
noise intensity for weak and strong cases.
\begin{table}[b!]
    \caption{\label{tab:Lorenz_par} Inference results for the parameters of the system
    (\ref{eq:Lorenz}) with weak (first set) and strong (second set) dynamical noise.  A synthetic
    data set of 4,000 points was generated for each case by simulating the system with a diffusion
    matrix $\hat{\bf D} = \tau^2 \, \hat{\bf I}$, and subsequently sampling its trajectory with $h =
    0.002$.  \\}
    \begin{center}
    \begin{tabular}{c @{\hspace{0.25in}} r @{.} l @{\hspace{0.45in}} r @{.} l}
        \hline \hline
        {\it Parameter} & \multicolumn{2}{l}{\it Value} & \multicolumn{2}{l}{\it Estimate} \\
        \hline
        $\sigma$ & $10$&$00$ & $9$&$9916$ \\
        $r$ & $28$&$00$ & $27$&$8675$ \\
        $b$ & $2$&$667$ & $2$&$6983$ \\
        $\tau$ & $1$&$00$ & $0$&$9965$ \\
        \hline
        $\sigma$ & $10$&$00$ & $9$&$9039$ \\
        $r$ & $28$&$00$ & $28$&$3004$ \\
        $b$ & $2$&$667$ & $2$&$8410$ \\
        $\tau$ & $40$&$00$ & $39$&$9108$ \\
        \hline \hline
    \end{tabular}
    \end{center}
\end{table}

Now we turn to an extension of our approach that allows for
efficient identification of the Lorenz system from a substantially
extended model space with 33 unknown parameters.

\subsubsection{Model reconstruction with strong dynamical noise}

When the analytical form of the nonlinear force field is not known
{\em a priori}, one may adopt a parametric model, as was done in
(\ref{eq:model}); in this setting, it is more appropriate to refer
to the inference problem as model reconstruction.  In practical
terms, the main difference between parameter estimation and model
reconstruction is in the number of unknown parameters involved,
which is typically an order of magnitude larger in the latter
case.  This proliferation of unknowns is one of the main reasons
why inference methods that rely on brute-force numerical
techniques are rendered largely impracticable for model
reconstruction.  On the other hand, owing to its analytical
foundation, our algorithm is quite capable of handling this more
difficult task, as we demonstrate below.

We start by considering the data set of Figure~\ref{fig:Lorenz},
where the structure of the Lorenz attractor is drastically
obscured by the presence of strong dynamical noise (almost 2
orders of magnitude stronger then in~\cite{Rossi:02a}). We wish to
fit this data set with a polynomial model of quadratic
nonlinearity.  Toward this end, we introduce a parametric model of
the form
\begin{equation}
   {\dot x}_{l} = \sum_{l' = 1}^{3} a_{l l'} \, x_{l'}(t) + \sum_{l', l'' = 1}^{3} b_{l l' l''} \,
   x_{l'}(t) \, x_{l''}(t) + \xi_{l}(t),
   \label{eq:Lorenz_mod}
\end{equation}
$l, l', l'' = 1, 2, 3$.  Including the elements of the (symmetric)
diffusion matrix $\hat{\bf D}$, we now have a total of 33 unknown
parameters comprising the set ${\cal M} = \{\{a_{l l'}\}, \{b_{l
l' l''}\}, \{D_{l l'}\}\}$.  Despite the restriction to linear,
bilinear, and quadratic polynomial basis functions,
(\ref{eq:Lorenz_mod}) still represents an extremely broad class of
dynamical models.  Assuming no measurement noise for simplicity,
the application of our algorithm entails the use of equations
(\ref{eq:updateD}) and (\ref{eq:updateC}) with (\ref{eq:defs_2})
and (\ref{eq:defs_3}).  The inferred parameter values are shown in
Table~\ref{tab:Lorenz_mod}; it can be seen that, even in this case
of extremely strong dynamical noise, our algorithm succeeds in
accurately reconstructing the Lorenz model.
\begin{table}
    \caption{\label{tab:Lorenz_mod} Inference results for the parameters of the model
    (\ref{eq:Lorenz_mod}).  (For brevity, only a representative subset of the $b_{l l' l''}$ and
    $D_{l l'}$ parameters is shown.)  Synthetic data, comprising 200 blocks of 600,000 points each,
    were generated by simulating the system with the standard Lorenz parameter set and a diagonal
    diffusion matrix, and subsequently sampling its trajectory with $h = 0.005$.  \\}
    \begin{center}
    \begin{tabular}{c @{\hspace{0.25in}} r @{.} l @{\hspace{0.35in}} r @{.} l}
        \hline \hline
        {\it Parameter} & \multicolumn{2}{l}{\it Value} & \multicolumn{2}{l}{\it Estimate} \\
        \hline
        $a_{11}$ & $-10$&$00$ & $-10$&$55$ \\
        $a_{21}$ & $28$&$00$ & $27$&$53$ \\
        $a_{31}$ & $0$&$0$ & $-0$&$43$ \\
        $a_{12}$ & $10$&$00$ & $10$&$77$ \\
        $a_{22}$ & $-1$&$00$ & $-0$&$194$ \\
        $a_{32}$ & $0$&$0$ & $0$&$596$ \\
        $a_{13}$ & $0$&$0$ & $0$&$065$ \\
        $a_{23}$ & $0$&$0$ & $0$&$001$ \\
        $a_{33}$ & $-2$&$667$ & $-2$&$759$ \\
        $b_{111}$ & $0$&$0$ & $0$&$013$ \\
        $b_{211}$ & $0$&$0$ & $0$&$001$ \\
        $b_{311}$ & $0$&$0$ & $0$&$018$ \\
        $b_{112}$ & $0$&$0$ & $0$&$002$ \\
        $b_{212}$ & $0$&$0$ & $-0$&$012$ \\
        $b_{312}$ & $1$&$00$ & $0$&$995$ \\
        $b_{113}$ & $0$&$0$ & $-0$&$016$ \\
        $b_{213}$ & $-1$&$00$ & $-0$&$985$ \\
        $D_{11}$ & $1500$&$0$ & $1522$&$1$ \\
        $D_{22}$ & $1600$&$0$ & $1621$&$5$ \\
        $D_{33}$ & $1700$&$0$ & $1713$&$4$ \\
        \hline \hline
    \end{tabular}
    \end{center}
\end{table}

\subsubsection{Accuracy of the inferred parameters}

\begin{figure}
    \centering
    \psfrag{m}[cr][cr]{$\langle a_{21} \rangle$}
    \psfrag{v}[cr][cr]{$\langle \Delta a_{21}^{2} \rangle$}
    \psfrag{T}[cc][cc]{$T$}
    \psfrag{(a)}[cc][cc]{(a)}
    \psfrag{(b)}[cc][cc]{(b)}
    \includegraphics[width=3.25in]{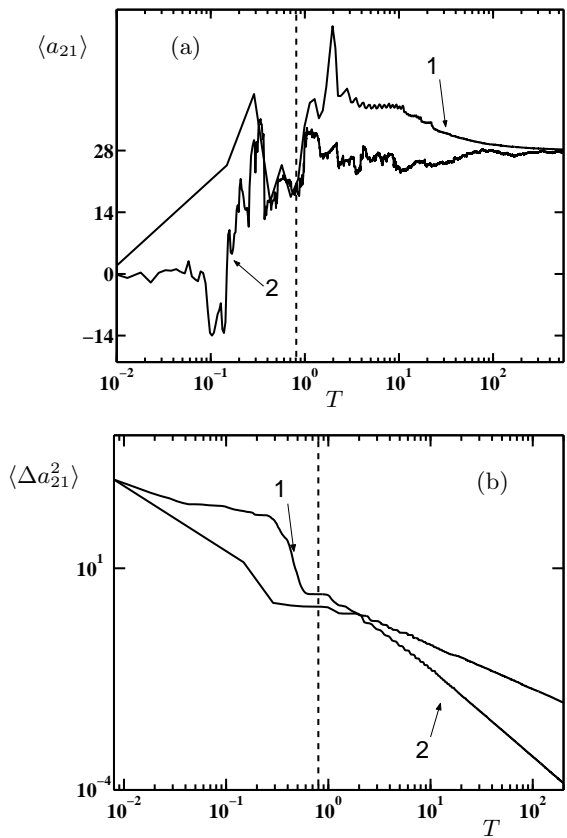}
    \caption{\label{fig:Lorenz_a} Results for (a) the posterior mean (i.e., inferred value) and (b)
    the posterior variance (i.e., associated uncertainty) of the model parameter $a_{21}$
    corresponding to parameter $r$ of the Lorenz system (\ref{eq:Lorenz_mod}) as a function of
    increasing observation duration.  Curve {\sf 1}: $\{D_{l l}\} = \{0.01, 0.012, 0.014\}$, $h =
    0.002$; curve {\sf 2}: $\{D_{l l}\} = \{100, 120, 140\}$, $h = 0.00002$.  The time instant of
    step-like decrease in the variance is indicated by the vertical dashed line.}
\end{figure}

The accuracy of the reconstruction depends on a number of factors.  As an example, consider the
inferred values and variances of the Lorenz parameter $r$ as a function of the total observation
duration, shown in Figure~\ref{fig:Lorenz_a} for two different levels of noise.
Of particular note is a sharp, step-like decrease in the variances that occurs on the same time
scale as the period of system oscillations, $\tau_{\rm{osc}} \simeq 0.6$ (marked by the dashed line
in Figure~\ref{fig:Lorenz_a}).  In addition to the total observation duration $T$, the inference
error is also sensitive to the values of the sampling interval $h$ and the noise intensities $D_{l
l}$.  For example, for the parameters of curve {\sf 1} in Figure~\ref{fig:Lorenz_a}, the relative
inference error was $0.015\%$.  When the noise intensity was increased by a factor of $10^4$ (curve
{\sf 2} in Figure~\ref{fig:Lorenz_a}), the ratio $T/h$ (i.e., the number of data points $N$) had to
be increased by at least a factor of $250$ to achieve an inference error below $1\%$.

We have observed that it is generally possible to achieve arbitrarily accurate inference results
with a (sufficiently small) fixed sampling interval by increasing the total duration of observation;
this is true even in the case of a full (i.e., non-diagonal) diffusion matrix.  Indeed, we were able
to achieve highly accurate parameter estimates for sampling intervals ranging from $10^{-6}$ to
$0.01$ and noise intensities ranging from $0$ to $10^{2}$.  As an example, we summarize in
Table~\ref{tab:Lorenz_full} our inference results for the model (\ref{eq:Lorenz_mod}) with a full
diffusion matrix, showing extremely high accuracy.
\begin{table}
    \caption{\label{tab:Lorenz_full} Inference results for the parameters of the model
    (\ref{eq:Lorenz_mod}), obtained using 200 blocks of 600,000 data points each, sampled at $h =
    0.005$.  True and inferred parameter values are shown along with the corresponding error
    (relative and absolute errors for the nonzero and zero parameters, respectively).  The inference
    error is below $1\%$ for all parameters, and much less for most.  \\}
    \begin{center}
    \begin{tabular}{c @{\hspace{0.25in}} r @{.} l @{\hspace{0.35in}} r @{.} l @{\hspace{0.5in}}
    r @{.} l}
        \hline \hline
        {\it Parameter} & \multicolumn{2}{l}{\it Value} & \multicolumn{2}{l}{\it Estimate} &
        \multicolumn{2}{l}{$\%$ {\it error}} \\
        \hline
        $a_{11}$ & $-10$&$0000$ & $-9$&$9984$ & $0$&$0161$ \\
        $a_{21}$ & $28$&$0000$ & $28$&$0139$ & $0$&$0496$ \\
        $a_{31}$ & $0$&$0$ & $-0$&$0052$ & $-0$&$5180$ \\
        $a_{21}$ & $10$&$0000$ & $9$&$9982$ & $0$&$0178$ \\
        $a_{22}$ & $-1$&$0000$ & $-1$&$0051$ & $0$&$5120$ \\
        $a_{23}$ & $0$&$0$ & $0$&$0031$ & $0$&$3072$ \\
        $a_{31}$ & $0$&$0$ & $0$&$0014$ & $0$&$1390$ \\
        $a_{32}$ & $0$&$0$ & $0$&$0015$ & $0$&$1542$ \\
        $a_{33}$ & $-2$&$6667$ & $-2$&$6661$ & $0$&$0196$ \\
        $b_{111}$ & $0$&$0$ & $0$&$0002$ & $0$&$0179$ \\
        $b_{211}$ & $0$&$0$ & $0$&$0002$ & $0$&$0238$ \\
        $b_{311}$ & $0$&$0$ & $-0$&$0004$ & $-0$&$0401$ \\
        $b_{112}$ & $0$&$0$ & $-0$&$0002$ & $-0$&$0208$ \\
        $b_{212}$ & $0$&$0$ & $-0$&$0002$ & $-0$&$0223$ \\
        $b_{312}$ & $1$&$0000$ & $1$&$0006$ & $0$&$0607$ \\
        $b_{113}$ & $0$&$0$ & $-0$&$0001$ & $-0$&$0111$ \\
        $b_{213}$ & $-1$&$0000$ & $-1$&$0004$ & $0$&$0446$ \\
        $D_{11}$ & $0$&$2867$ & $0$&$2865$ & $0$&$0587$ \\
        $D_{22}$ & $0$&$4087$ & $0$&$4081$ & $0$&$1564$ \\
        $D_{33}$ & $0$&$5118$ & $0$&$5148$ & $0$&$5946$ \\
        $D_{12} = D_{21}$ & $0$&$2052$ & $0$&$2049$ & $0$&$1442$ \\
        $D_{13} = D_{31}$ & $0$&$1069$ & $0$&$1061$ & $0$&$7657$ \\
        $D_{23} = D_{32}$ & $0$&$1814$ & $0$&$1812$ & $0$&$1028$ \\
        \hline \hline
    \end{tabular}
    \end{center}
\end{table}

\subsubsection{Optimal compensation for the noise-induced errors}
\begin{figure}
    \centering
    \psfrag{coef}[cc][cc]{$r$}
    \psfrag{hist}[cc][cc]{$p(r)$}
    \includegraphics[width=3.25in]{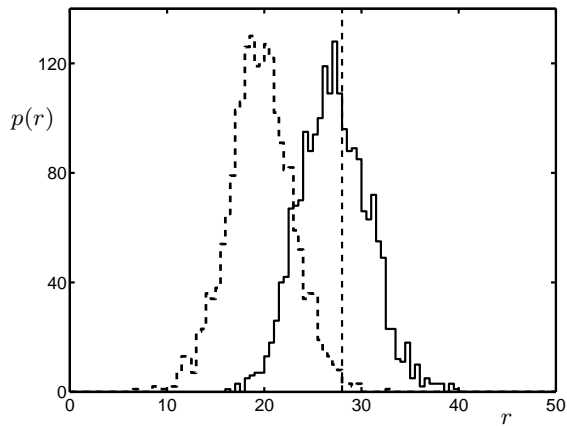}
    \caption{\label{fig:Lorenz_b} Demonstration of improved inference accuracy due to the prefactor
    (\ref{eq:prefactor}) in the likelihood function.  The true value of the parameter being inferred
    is indicated by the vertical dashed line.  The solid and dashed curves respectively show the
    histograms of parameter values inferred by our algorithm and by the generalized least-squares
    method, which lacks the Jacobian prefactor.  The histograms were built from an ensemble of 1,000
    numerical experiments with 90,000 data points each.}
\end{figure}

Finally, we would like to demonstrate the importance of the
Jacobian prefactor (\ref{eq:prefactor}) included in our likelihood
function by examining the inference results obtained with and
without this term.  As shown in Figure~\ref{fig:Lorenz_b} for
parameter $r$ of the Lorenz system, the omission of the prefactor
in the likelihood function results in a systematic underestimation
of this parameter, whereas the inclusion of this term leads to an
accurate inference as it optimally compensates for the effects of
dynamical noise.

\subsubsection{Discussion of results}

The Lorenz system provides a concrete example with which to
emphasize the advantages of our algorithm over previous work.  We
note firstly that we derive the correct form of the likelihood
function and  avoid using of {\it ad hoc} likelihood function
introduced in~\cite{Rossi:02a}. Furthermore, we obtain analytical
solution of the problem. This innovations allow us to estimate
model parameters of the Lorenz system much faster and more
accurately using the same number of points as in~\cite{Rossi:02a}.
Furthermore, our results unlike the results reported
in~\cite{Rossi:02a} do not depend neither on the choice of initial
values for the model parameters nor on the {\it ad hoc} conditions
imposed on the analysis of the experimental data to exclude points
from certain regions of the phase space.

The computational efficiency of our algorithm also allows us to
lift the practical limitation on the total number of data points
used for inference in previous work. (The relatively small number
of points (4,000) used for inference in~\cite{Rossi:02a} was
dictated by the complexity of the extensive numerical optimization
algorithms used therein.)  In our approach to inference,
processing of ${\cal O}(10^5)$ data points takes only a few
seconds on a personal computer with a 1-GHz CPU, therefore
enabling the use of very large data sets to achieve arbitrarily
accurate model reconstruction.

More importantly, the efficiency of our algorithm allows us to
extend substantially the dimensionality of the model space. As a
consequence it can  be efficiently applied to deal with a more
general problem of model reconstruction, when the functional form
of a nonlinear vector field is unknown

\subsection{A system of five coupled oscillators}
\label{ss:five_coupled}
The limitations of inference algorithms that rely on numerical methods for global optimization and
multi-dimensional integration come into sharper focus when systems with large numbers of model
parameters are investigated.  We now wish to illustrate the advantages of our algorithm by inferring
a model for a system comprising five locally- and globally-coupled van der Pol oscillators with
${\cal O}(10^{2})$ unknown model parameters.

With $K = 5$, the system under study is
\begin{equation}
    \label{eq:5vdp}
    \left.
    \hspace{-0.15in}
    \begin{array}{rcl}
    {\dot u}_k & = & v_k, \\ \\
    {\dot v}_k & = & \varepsilon_l \, (1 - u_k^2) \, v_k - \omega_l \, u_k + \sum_{k' = 1
    \setminus k}^K \eta_{k k'} \, u_{k'} \\ \\
    & + & u_k \left[\gamma_{k(k-1)} \, u_{k-1} + \gamma_{k(k+1)} \, u_{k+1}\right] \\ \\
    & + & \sum_{k' = 1}^K \sigma_{k k'} \, \mu_{k'},
    \end{array}
    \ \right\}
\end{equation}
where $\{\mu_{k}(t)\}$ are mutually independent, zero-mean,
unit-variance, delta-correlated, Gaussian noise processes.  We
assume (for simplicity) that there is no measurement noise, and
that the state is partially observed to produce the signal ${\bf
y} = [v_1 \ v_2 \ v_3 \ v_4 \ v_5]^{\rm T}$. The state of the
system is thus described by the 10-dimensional vector ${\bf x} =
[u_1 \ \ldots \ u_5 \ v_1 \ \ldots \ v_5]^{\rm T}$. We note,
however, that values of $u_k$  in the model (\ref{eq:5vdp}) are
assumed to be know and do not have to be inferred. Therefore the
problem is reduced to the inference of the model parameters of
five couple equations for $v_k$, which are parameterized according
to the (\ref{eq:model}) with $\hat{\bf D} = \sigma \sigma^T$.

\begin{figure}
    \centering
    \psfrag{x1}[cc][cc]{$u_{1}$}
    \psfrag{x2}[cc][cc]{$u_{2}$}
    \psfrag{x3}[cc][cc]{$u_{3}$}
    \psfrag{(a)}[cc][cc]{(a)}
    \psfrag{(b)}[cc][cc]{(b)}
    \includegraphics[width=3.25in]{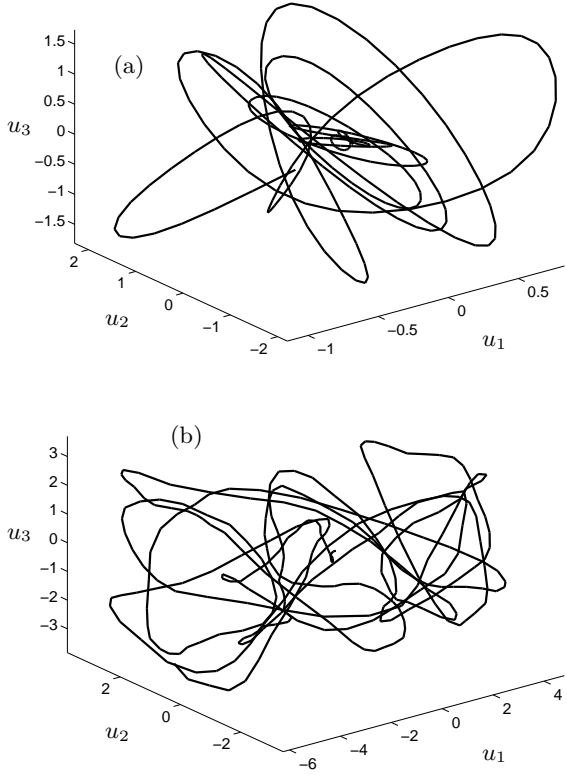}
    \caption{\label{fig:vanderpol} The phase portrait of the system (\ref{eq:5vdp}) as projected
    onto the $(u_1, u_2, u_3)$ subspace: (a)~deterministic system; (b)~stochastic system with a
    diagonal diffusion matrix of the form $\hat{\bf D} = 100 \, \hat{\bf I}$.  (Note the scale
    change between the axes of the two figures.)  See Table~\ref{tab:1st_osc} and
    Figure~\ref{fig:coeff_5_b} for the values of some of the model parameters used in the
    simulation.}
\end{figure}

The phase portrait of this system, projected onto the $(u_1, u_2,
u_3)$ subspace of its phase space, is shown in
Figure~\ref{fig:vanderpol} for some nominal set of model
parameters.

\begin{figure}
    \psfrag{coef}[cc][cc]{$a_{5 l'}$}
    \psfrag{diff}[cc][cc]{$Q_{5 l'}$}
    \psfrag{l!}[cl][cl]{\scriptsize{$Q_{5 1}$}}
    \psfrag{l\@}[cl][cl]{\scriptsize{$Q_{5 2}$}}
    \psfrag{l#}[cl][cl]{\scriptsize{$Q_{5 3}$}}
    \psfrag{l\$}[cl][cl]{\scriptsize{$Q_{5 4}$}}
    \psfrag{l\%}[cl][cl]{\scriptsize{$Q_{5 5}$}}
    \psfrag{l\^}[cl][cl]{\scriptsize{$\varepsilon_{5}$}}
    \psfrag{l\&}[cl][cl]{\scriptsize{$\omega_{5}$}}
    \psfrag{l*}[cl][cl]{\scriptsize{$\eta_{5 1}$}}
    \psfrag{l~}[cl][cl]{\scriptsize{$\eta_{5 2}$}}
    \psfrag{l-}[cl][cl]{\scriptsize{$\eta_{5 3}$}}
    \psfrag{l+}[cl][cl]{\scriptsize{$\eta_{5 4}$}}
    \psfrag{(a)}[cc][cc]{(a)}
    \psfrag{(b)}[cc][cc]{(b)}
    \includegraphics[width=3.25in]{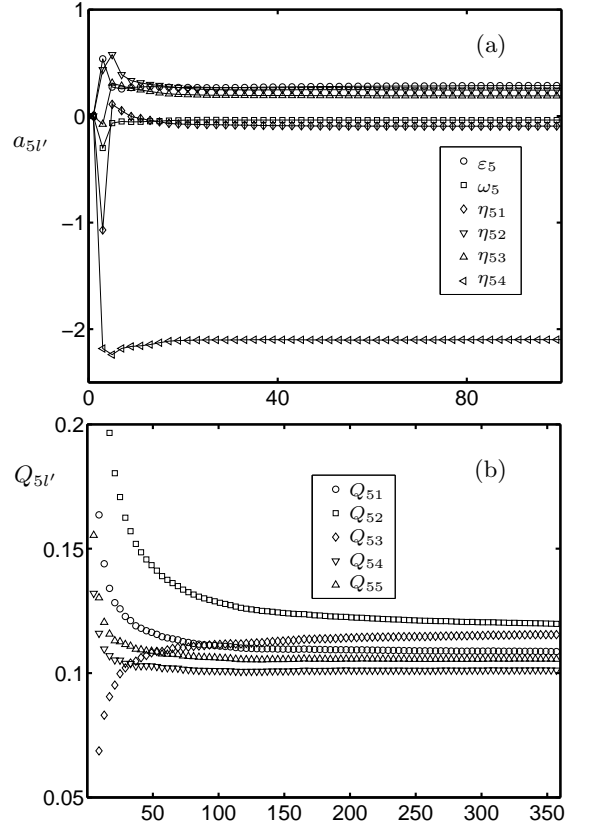}
    \caption{\label{fig:coeff_5_b} Accurate inference of (a) the parameters of the fifth oscillator
    in the system (\ref{eq:5vdp}) and (b) the elements of the last row of the diffusion matrix
    $\hat{\bf Q}$.  The horizontal axes show the number of blocks of data used, with 800 points in
    each block, sampled at $h = 0.02$.}
\end{figure}

We choose the following basis functions with which to reconstruct the model:
\begin{eqnarray*}
    \phi_k & = & u_k, \\
    \phi_{k + K} & = & v_k, \\
    \phi_{k + 2 K} & = & u_k^2 \, v_k, \\
    \phi_{1 + 3 K} & = & u_{1}^2, \\
    \phi_{2 + 3 K} & = & u_1 \, u_2, \\
    & \vdots & \\
    \phi_{15 + 3 K} & = & u_5^2,
\end{eqnarray*}
$k = 1, 2, \ldots, K$.  Together with the elements of the
(symmetric) diffusion matrix $\hat{\bf D}$, we thus have a total
of 165 model parameters to infer.  We summarize in
Table~\ref{tab:1st_osc} the results of our algorithm for the first
oscillator, once again showing high inference accuracy.
Additionally, the convergence of the parameters of the fifth
oscillator and the noise intensities to their correct values is
shown in Figure~\ref{fig:coeff_5_b} as a function of the amount of
data used.
\begin{table}
    \caption{\label{tab:1st_osc} Inference results for the parameters of the first oscillator in the
    system (\ref{eq:5vdp}), obtained using 50 blocks of 150,000 data points each, sampled at $h =
    0.06$.  The inference error is well below 1\% for all parameters.  \\}
    \begin{center}
    \begin{tabular}{c @{\hspace{0.25in}} r @{.} l @{\hspace{0.35in}} r @{.} l @{\hspace{0.5in}}
    r @{.} l}
        \hline \hline
        {\it Parameter} & \multicolumn{2}{l}{\it Value} & \multicolumn{2}{l}{\it Estimate} &
        \multicolumn{2}{l}{$\%$ {\it error}} \\
        \hline
        $\varepsilon_1$ & $-8$&$40$ & $-8$&$4167$ & $0$&$2$ \\
        $\omega_1$ & $-4$&$4000$ & $-4$&$4031$ & $0$&$07$ \\
        $\eta_{12}$ & $0$&$4400$ & $0$&$4432$ & $0$&$7$ \\
        $\eta_{13}$ & $-0$&$60$ & $-0$&$6033$ & $0$&$54$ \\
        $\eta_{14}$ & $0$&$96$ & $0$&$9625$ & $0$&$3$ \\
        $\eta_{15}$ & $0$&$80$ & $0$&$8022$ & $0$&$3$ \\
        $\gamma_{12}$ & $-0$&$480$ & $-0$&$4806$ & $0$&$1$ \\
        $\gamma_{15}$ & $0$&$8$ & $0$&$8013$ & $0$&$2$ \\
        $Q_{11}$ & $0$&$20$ & $0$&$2020$ & $1$&$0$ \\
        \hline \hline
    \end{tabular}
    \end{center}
\end{table}

\begin{figure}
    \centering
    \psfrag{coef}[cc][cc]{$\varepsilon_{1}$}
    \psfrag{hist}[cc][cc]{$p(\varepsilon_{1})$}
    \includegraphics[width=3.25in]{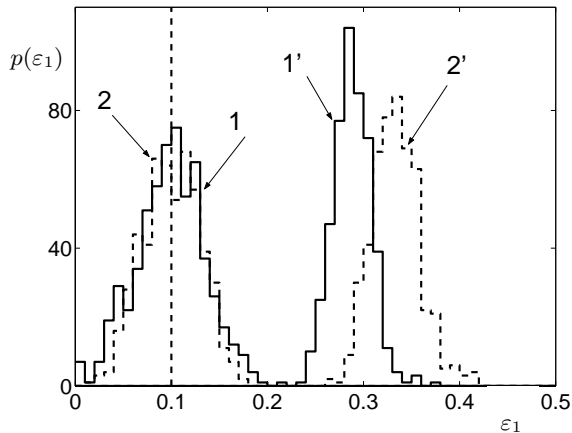}
    \caption{\label{fig:compensation} Further demonstration of improved inference accuracy due to
    the prefactor (\ref{eq:prefactor}) in the likelihood function.  The true value of the parameter
    being inferred is indicated by the vertical dashed line.  Histograms {\sf 1} and {\sf 2} show
    results obtained with our algorithm, while histograms {\sf 1'} and {\sf 2'} are due to the GLS
    method, showing the detrimental effect on inference accuracy of the missing prefactor.  The
    diffusion matrix was $\hat{\bf Q}$ for curves {\sf 1} and {\sf 1'}, and $2 \, \hat{\bf Q}$ for
    curves {\sf 2} and {\sf 2'} (see text).}
\end{figure}

In order to further highlight the vital noise compensation effect
provided by the prefactor term in the likelihood function used in
the present work, we compare in Figure~\ref{fig:compensation}
inference results for one of the coefficients of the system
(\ref{eq:5vdp}), $\varepsilon_1$, obtained with two different
diffusion matrices $\hat{\bf D}$ and $\hat{\bf D}/4$, where the
matrix $\hat{\bf D}$ chosen at random to be
\begin{equation}
    \label{eq:matrix}
    \hat{\bf D} = \left[
    \begin{array}{ccccc}
    $3.9628$ & $2.9636$ & $0.6176$ & $2.4941$ & $2.5068$ \\
    $2.9636$ & $5.5045$ & $2.7690$ & $5.2893$ & $5.5421$ \\
    $0.6176$ & $2.7690$ & $4.6974$ & $4.8813$ & $3.0284$ \\
    $2.4941$ & $5.2893$ & $4.8813$ & $7.1428$ & $4.6732$ \\
    $2.5068$ & $5.5421$ & $3.0284$ & $4.6732$ & $7.5784$ \\
    \end{array}
    \right].
\end{equation}
As discussed earlier, without the Jacobian prefactor (\ref{eq:prefactor}), (\ref{eq:updateC})
reduces to the GLS estimator.  Figure~\ref{fig:compensation} shows that the GLS estimator
systematically overestimates the value of $\varepsilon_1$; the larger the noise intensity, the
larger the systematic error, reaching a few hundred per cent in this case, as shown by curves {\sf
1'} and {\sf 2'}.  On the other hand, when the proper Jacobian prefactor is included in the
likelihood function as in (\ref{eq:likelihood}), we are able to achieve optimal compensation of the
noise-induced errors, as shown by curves {\sf 1} and {\sf 2} obtained with the same noise
intensities.

\section{Discussion}
\label{s:discussion}
In this paper, we introduced a novel technique for inferring the unknown parameters of stochastic
nonlinear dynamical systems from time-series data.  The key features of our approach are
\begin{itemize}
    \item a likelihood function written in the form of a path integral over stochastic system
    trajectories, properly accounting for measurement noise and optimally compensating for dynamical
    noise; and

    \item a parameterization of the unknown force field that renders the inference problem
    essentially linear, despite the strong nonlinearity of the model itself.
\end{itemize}
Specifically, our analytical derivation produces the correct Jacobian prefactor in the likelihood
function, which was missed in earlier works.  Meanwhile, the representation of the system
nonlinearity as an expansion over a set of basis functions provides stable and robust inference for
a broad range of dynamical models.  These features enabled us to devise a highly accurate and
efficient Bayesian inference algorithm that can reconstruct models of stochastic nonlinear dynamical
systems {\em without} resorting to brute-force numerical optimization.

We illustrated the advantages of our approach by applying it first
to the inference of the stochastic nonlinear dynamical system of
Lorenz.  In the context of parameter estimation with 8 unknown
parameters, we showed that the accuracy and efficiency of our
algorithm exceed those achieved (under similar conditions) in
earlier works.  We also demonstrated that our algorithm can deal
with the Lorenz system in the more general setting of model
reconstruction, i.e., assuming no knowledge of the functional form
of the nonlinear vector field.  Although a much larger number of
33 unknown parameters were involved here, our algorithm was still
able to achieve a high inference accuracy.

In order to further illustrate the strengths of our algorithm, we
applied it next to a system of five coupled nonlinear noisy
oscillators.  Using a set of polynomial basis functions for the
nonlinear field and a full covariance matrix for the dynamical
noise, the model comprised 165 unknown parameters, all of which
were inferred within an error of $1\%$ from a data set of $10^5$
points, taking only a few seconds on a personal computer of
average computing power.  These demonstrations, we believe, are
convincing representation of the capability of our approach, in
both accuracy and efficiency, for reconstructing models of
stochastic nonlinear dynamical systems.

Furthermore, the efficiency of our algorithm has enabled us
recently to identify a stochastic nonlinear model of coupled
cardiovascular oscillators using univariate physiological times
series data~\cite{Smelyanskiy:05} thus opening a new venue for a
broad range of important interdisciplinary applications.

Several simplifying assumptions were made here to provide a clear description of the algorithm in
its barest form.  Although the examples of Section~\ref{s:models} dealt with noiseless measurements,
we have indicated in Section~\ref{s:main} how measurement noise can be included systematically in
our inference algorithm.  These examples also required the use of polynomials only, but the approach
is in fact completely flexible regarding the type of basis functions used to model the nonlinear
force, including time delays.  Additionally, the path-integral technique used here to derive the
likelihood function allows for a number of straightforward generalizations of our algorithm to the
reconstruction of models with colored and multiplicative (or parametric) dynamical noise, and
arbitrary (i.e., not necessarily uniform or short) sampling intervals.  Finally, although the basic
theory of Section~\ref{s:main} was developed under the implicit assumption that all dynamical
variables are available for measurement, we have shown in Section~\ref{ss:five_coupled} that our
algorithm is able to reconstruct a complete model from partial measurements of the system
trajectory.  Since it is often not feasible to measure all degrees of freedom in practice, a
generalization of our algorithm to deal with ``hidden variables'' will be very useful.  These
extensions will be explored in subsequent publications.

\section*{Acknowledgment}
This work was supported by the Engineering and Physical Sciences Research Council (UK), NASA
CICT--IS--IDU Project (USA), the Russian Foundation for Fundamental Science, and INTAS.

\section*{Appendix A \\ Maximum-likelihood parameter estimation \\ for a one-dimensional system}
 \label{appendixA}

Consider a one-dimensional stochastic nonlinear dynamical system
\begin{eqnarray}
    \label{eq:1d}
    {\dot x}(t) = f(x) + \xi(t),
\end{eqnarray}
where $\xi(t)$ is a zero-mean Gaussian noise process with $\langle\xi(t) \, \xi(0)\rangle = D \,
\delta(t)$.  The mid-point approximation to (\ref{eq:1d}) on the time lattice $\{t_n = t_{0} + n h,
n = 0, 1, \ldots, N\}$ is
\begin{eqnarray}
    \label{eq:1d_discrete}
    x_{n+1} = x_n + h \, f({\tilde x}_n) + z_n,
\end{eqnarray}
where we used $x_{n} \equiv x(t_{n})$ and ${\tilde x}_n \equiv \frac{1}{2} \, (x_{n+1} + x_{n})$,
and $z_n \equiv \int_{t_{n}}^{t_{n} + h} \xi(t) \, {\rm d}t$ form a sequence of zero-mean Gaussian
random variables with $\langle z_n \, z_{n'}\rangle = h \, D \, \delta_{n n'}$.

The probability of realization of a particular random sequence $\{z_n\}$ is
\begin{eqnarray}
    \label{eq:1d_random_force}
    {\cal P}[\{z_n\}] = \prod_{n=0}^{N-1}\frac{{\rm d}z_{n}}{\sqrt{2 \pi h D}} \, \exp
    \left(-\frac{z_n^2}{2 h D}\right).
\end{eqnarray}
Using the Markovian property of $x(t)$ and the transformation rule $p(\{z_n\}) \, \prod_{n} {\rm
d}z_n = p(\{x_n\}) \, \prod_{n} {\rm d}x_n$, along with (\ref{eq:1d_discrete}) and
(\ref{eq:1d_random_force}), we find the probability {\em density} of the dynamical system trajectory
to be
\begin{eqnarray}
    \label{eq:1d_random_trajectory}
    p(\{x_n\}) & = & p_{\rm st}(x_0) \, J(\{x_n\}) \, (2 \pi h D)^{-N/2} \nonumber \\
    & \times & \prod_{n=0}^{N-1} \exp \left\{-\frac{h}{2 D} \, [{\dot x}_n - f({\tilde
    x}_n)]^2\right\},
\end{eqnarray}
where ${\dot x}_n \equiv (x_{n+1} - x_{n})/h$, and the Jacobian of the transformation, to lowest
order in $h$, is
\begin{equation*}
    J(\{x_n\}) \simeq \prod_{n=0}^{N-1} \left[1 - \frac{h}{2} \, f'({\tilde x}_n)\right] \simeq
    \exp \left[-\frac{h}{2} \, \sum_{n=0}^{N-1} f'({\tilde x}_n)\right].
\end{equation*}
Here, prime indicates differentiation with respect to argument.  Thus, we obtain for the
negative-logarithm of (\ref{eq:1d_random_trajectory}) the expression
\begin{eqnarray}
    \label{eq:1d_cost}
    S & = & -\ln p_{\rm st}(x_0) + \frac{N}{2} \, \ln (2 \pi h D) \nonumber \\
    & + & \frac{h}{2} \, \sum_{n=0}^{N-1} \left\{ f'({\tilde x}_n) + \frac{1}{D} \, [{\dot x}_n -
    f({\tilde x}_n)]^2\right\}.
\end{eqnarray}

Assume now that we observe a time series $\{x_n\}$, and wish to reconstruct a one-dimensional
stochastic nonlinear dynamical model for the system that generated the data; i.e., infer the form of
the nonlinear function $f(x)$ and estimate the noise intensity $D$ in (\ref{eq:1d}).  A fruitful
approach to this problem is to model the nonlinearity as a {\em linear} superposition of a set of
nonlinear basis functions:
\begin{equation}
    \label{eq:1d_model}
    f(x) = \sum_{m=1}^{M} c_{m} \, u_{m}(x) = {\bf c}^{\rm T} \, {\bf u}(x).
\end{equation}
The maximum-likelihood (ML) estimates for the unknown model parameters $\bf c$ and $D$ are then
furnished by the global minimum of $S$.  Thus, setting $\partial S/\partial D = 0$ and passing to
the limit $h \to 0$ with $T = N h$, we find
\begin{equation}
    \label{eq:1d_noise}
   D = \frac{1}{N} \, \int_{t_{0}}^{t_{0} + T} \left[\dot x - {\bf c}^{\rm T} \, {\bf u}(x)\right]^2
   {\rm d}t.
\end{equation}
Next, substituting (\ref{eq:1d_model}) into (\ref{eq:1d_cost}) and rearranging, we obtain $S = \rho
- {\bf c}^{\rm T} \, {\bf w} + \frac{1}{2} \, {\bf c}^{\rm T} \, \hat{\boldsymbol \Xi} \, {\bf c}$,
where
\begin{eqnarray*}
   \label{eq:1d_parameters}
   \rho & = & -\ln p_{\rm st}(x_0) + \frac{N}{2} \, \ln(2 \pi h D) + \frac{1}{2 D} \,
   \int_{t_{0}}^{t_{0} + T} {\dot x}^2 \, {\rm d}t, \\
   {\bf w} & = & \int_{t_{0}}^{t_{0} + T} \left[\frac{1}{D} \, {\dot x} \, {\bf u}(x) - \frac{1}{2}
   \, \frac{\partial {\bf u}(x)}{\partial x}\right] {\rm d}t, \\
   \hat{\boldsymbol \Xi} & = & \frac{1}{D} \, \int_{t_{0}}^{t_{0} + T} {\bf u}(x) \, {\bf u}^{\rm
   T}(x) \, {\rm d}t.
\end{eqnarray*}
The condition $\partial S/\partial {\bf c} = 0$ now gives
\begin{equation}
   \label{eq:1d_coef}
   {\bf c} = \hat{\boldsymbol \Xi}^{-1} \, {\bf w}.
\end{equation}
The ML estimates are found by iterating (\ref{eq:1d_noise}) and (\ref{eq:1d_coef}) to convergence.

In Section~\ref{s:main}, this theory is extended to deal with multi-dimensional system models and to
include prior information on model parameters; it is particularly interesting to contrast the
results above with our main algorithm given in Section~\ref{ss:algorithm}.

\section*{Appendix B \\ The generalized least-squares estimator}
 \label{appendixB}

It is insightful to contrast the algorithm presented in this paper
with the generalized least-squares (GLS) estimator.  Starting
again with the system (\ref{eq:dynamics}), we neglect measurement
noise, adopt the parameterization of (\ref{eq:model}), and apply
the mid-point approximation, obtaining
\begin{equation}
    \dot{\bf y}_n = {\hat {\bf U}}_n \, {\bf c} + {\boldsymbol \zeta}_n, \quad n = 0, 1, \ldots,
    N-1,
    \label{eq:partitioned_dynamics}
\end{equation}
where, as before, we introduced $\dot{\bf y}_n = ({\bf y}_{n+1} - {\bf y}_n)/h$ and $\hat{\bf U}_n =
\hat{\bf U}(\tilde{\bf y}_{n})$ with $\tilde{\bf y}_{n} = \frac{1}{2} \, ({\bf y}_{n+1} + {\bf
y}_n)$.  The vectors $\{{\boldsymbol \zeta}_n\}$ satisfy
\begin{equation*}
    \langle {\boldsymbol \zeta}_{n} \rangle = {\bf 0}, \quad \langle {\boldsymbol \zeta}_{n} \,
    {\boldsymbol \zeta}_{n'}^{\rm T} \rangle = \frac{1}{h} \, \hat{\bf D} \, \delta_{n n'}.
\end{equation*}

We may arrange the $N$ equations contained in (\ref{eq:partitioned_dynamics}) into a single
partitioned matrix equation as
\begin{equation}
    \label{eq:partitioned_form}
    {\bf d} = \hat{\bf H} \, {\boldsymbol \gamma} + {\bf n},
\end{equation}
where
\begin{equation*}
    \hat{\bf H} = \left[
    \begin{array}{cccc}
    \hat{\bf U}_0 & \hat{\bf 0} & \ldots & \hat{\bf 0} \\
    \hat{\bf 0} & \hat{\bf U}_1 & \ldots & \hat{\bf 0} \\
    \vdots & \vdots & \ddots & \vdots \\
    \hat{\bf 0} & \hat{\bf 0} & \ldots & \hat{\bf U}_{N-1}
    \end{array} \right],
\end{equation*}
$\boldsymbol \gamma$ is a column vector comprising $N$ copies of the unknown model coefficient
vector $\bf c$, and ${\bf d} = \left[\dot{\bf y}_0 \ \dot{\bf y}_1 \ \ldots \ \dot{\bf
y}_{N-1}\right]^{\rm T}$ and ${\bf n} = \left[{\boldsymbol \zeta}_0 \ {\boldsymbol \zeta}_1 \ \ldots
\ {\boldsymbol \zeta}_{N-1}\right]^{\rm T}$ are composite data and noise vectors, respectively, the
latter having zero mean and a covariance matrix of the form
\begin{equation*}
    \langle {\bf n} \, {\bf n}^{\rm T} \rangle = \hat{\boldsymbol \Lambda} = \frac{1}{h} \left[
    \begin{array}{cccc}
    \hat{\bf D} & \hat{\bf 0} & \ldots & \hat{\bf 0}\\
    \hat{\bf 0} & \hat{\bf D} & \ldots & \hat{\bf 0} \\
    \vdots & \vdots & \ddots & \vdots \\
    \hat{\bf 0} & \hat{\bf 0} & \ldots & \hat{\bf D}
     \end{array} \right].
\end{equation*}

Now, the GLS estimator for the vector $\boldsymbol \gamma$ in (\ref{eq:partitioned_form}) is given
by (see, e.g.,~\cite{Theil:83})
\begin{equation}
    \label{eq:partitioned_estimator}
    {\boldsymbol \gamma} = \left(\hat{\bf H}^{\rm T} \, \hat{\boldsymbol \Lambda}^{-1} \, \hat{\bf
    H}\right)^{-1} \, \hat{\bf H}^{\rm T} \, \hat{\boldsymbol \Lambda}^{-1} \, {\bf d}.
\end{equation}
Using the diagonal forms of the matrices $\hat{\bf H}$ and $\hat{\boldsymbol \Lambda}$, we can
extract from (\ref{eq:partitioned_estimator}) the following estimate for our model coefficient
vector:
\begin{equation}
    \label{eq:GLS_estimator}
    {\bf c} = \left( \sum_{n = 0}^{N - 1} \hat{\bf U}_{n}^{\rm T} \, \hat{\bf D}^{-1} \, \hat{\bf
    U}_{n} \right)^{-1} \, \sum_{n = 0}^{N - 1} \hat{\bf U}_{n}^{\rm T} \, \hat{\bf D}^{-1} \,
    \dot{\bf y}_{n}.
\end{equation}

A comparison of (\ref{eq:GLS_estimator}) with our corresponding result (\ref{eq:updateC}) is
facilitated by an examination of the definitions (\ref{eq:defs_2}) and (\ref{eq:defs_3}), whereupon
it is seen that, in the absence of prior information (i.e., $\hat{\boldsymbol \Sigma}_{\rm{pr}} \to
\hat{\bf 0}$), the only difference between the two estimates is the additional term $\frac{1}{2} \,
{\bf v}_{n}$ in our expression.  The importance of this extra term is borne out by the examples
given in Section~\ref{s:models}, where it is observed that the GLS estimator leads consistently to
grossly inaccurate parameter estimates, while our algorithm succeeds in achieving arbitrarily high
inference accuracy.

\end{document}